\documentclass[conference,letterpaper]{IEEEtran}

\usepackage[utf8]{inputenc} 
\usepackage[T1]{fontenc}
\usepackage{url}
\usepackage{ifthen,amssymb,graphicx}
\usepackage{cite,xcolor,stmaryrd}
\usepackage{bbm}
\usepackage[cmex10]{amsmath} 

\newtheorem{lemma}{Lemma}
\newtheorem{theorem}{Theorem}
\newtheorem{corollary}{Corollary}
\newtheorem{remark}{Remark}

\usepackage{soul}

\begin{document}
\title{Group Complete-$\{s\}$ Pliable Index Coding}


\author{Badri N.\ Vellambi, Sina Eghbal, Lawrence Ong, and Parastoo Sadeghi}

\author{\IEEEauthorblockN{Sina Eghbal, Badri N.\ Vellambi}
\IEEEauthorblockA{University of Cincinnati\\
eghbalsa@mail.uc.edu badri.vellambi@uc.edu}
\and
\IEEEauthorblockN{Lawrence Ong}
\IEEEauthorblockA{University of Newcastle\\
lawrence.ong@newcastle.edu.au}

\and

\IEEEauthorblockN{Parastoo Sadeghi} 
\IEEEauthorblockA{University of New South Wales (Canberra) \\
p.sadeghi@unsw.edu.au}
}


\maketitle


\begin{abstract}
This paper introduces a novel class of PICOD($t$) problems referred to as $g$-group complete-$S$ PICOD($t$) problems. It constructs a multi-stage achievability scheme to generate pliable index codes for group complete PICOD problems when $S = \{s\}$ is a singleton set. Using the maximum acyclic induced subgraph bound, lower bounds on the broadcast rate are derived for singleton $S$, which establishes the optimality of the achievability scheme for a range of values for $t$ and for any $g$ and $s$. For all other values, it is shown that the achievability scheme is optimal among the restricted class of  broadcast codes. 
\end{abstract}

\section{Introduction}

Index Coding~\cite{bar2011index} is a multicast problem where a single server conveys messages to several receivers over a noiseless broadcast channel. The problem is studied under the premise that each receiver has a subset of messages as side information and demands a specific message that is not in its side information. In a flexible variant of the problem called \emph{pliable} index coding~\cite{brahma2015pliable} (abbreviated usually as PICOD($t$)), receivers do not demand specific messages. Instead, each receiver is satisfied upon decoding \textit{any} $t$ messages not in its side information. Both problem formulations seek the shortest normalized broadcast length and a corresponding optimal code.

Finding optimal linear codes for index coding and PICOD($t$) problems is NP-hard~\cite{bar2011index, brahma2015pliable}. Despite this pessimistic result, index coding remains an active research area with results in two broad clusters: the design of index codes, and establishing fundamental limits on optimal index code length.

An index-coding problem is completely characterized by the side information of the receivers and the message(s) they demand. A problem can thus be captured by a directed graph where the nodes represent the message indices. Using this graph representation, index codes have been designed based on the graph components (cycles~\cite{chaudhry11}, cliques~\cite{birkkol2006}, partial cliques~\cite{birkkol2006}, interlinked cycles~\cite{thapaongjohnson17it}) or its coloring~\cite{shanmugamdimakislangberg13}. Coloring of confusion graphs, where each node represents a realization of the messages, has also been used to design codes~\cite{birkkol2006,thapaongjohnson19}; it must be remarked that the size of confusion graphs grows exponentially with the message size. Besides graph-based approaches, codes based on random-coding arguments have also been devised~\cite{arbabjolfaei13}. In~\cite{sharififar2022optimality}, matroid theory was used to show impossibility of optimal linear codes over fields with characteristic three, and optimal non-linear codes for small instances of the problem were explicitly constructed.

The graph representation of an index-coding problem allows us to not only design codes, but also converse results. Specifically, the size of a maximum acyclic induced subgraph (MAIS) is a lower bound to the optimal index code length~\cite{bar2011index}. Another approach to deriving converse results uses Shannon-type information inequalities, and this is referred to as the polymatroid lower bound~\cite{arbabjolfaei13}.  

Designing codes for the PICOD problem, on the other hand, took very different approaches: the code construction is mostly algorithmic. Two algorithmic coding schemes with proven bounds are as follows: Brahma and Fragouli~\cite{brahma2015pliable} proposed a randomized coding scheme which constructs codes of length $O(\min\{t\log^2{n}, t\log{n} + \log^3{n}\})$  if $m = O(n^\delta)$, where $n$ is the number of receivers and $m$ is the number of messages; Song et al.~\cite{song2017polynomial} utilized a greedy approach to construct codes whose length is bounded from above by $O(t \log{n} + \log^2{n})$. 

Further, algorithmic coding schemes without established theoretical upper bounds have been developed. Such algorithms typically produce codes of shorter average lengths than those produced by the previously mentioned algorithms. For instance, the Greedy Cover algorithm~\cite{brahma2015pliable} constructs subsets of messages (that are jointly coded) by greedily and sequentially selecting a message that maximizes its utility. An improved variation of this algorithm was recently proposed  in~\cite{eghbal2023improved}. Krishnan et al.~\cite{krishnan2024pliable} used a graph-based method to construct codes for PICOD($t$) problems. Their coding scheme involves a $t$-fold conflict-free coloring of the vertices of a hypergraph associated with the PICOD($t$).

Lower bounds for PICOD(1) were derived based on receivers that are \emph{absent} in the problems~\cite{ongvellambikliewer2019}. Although this approach builds on the concept of MAIS lower bound for index coding, the result is expressed in terms of how a side-information set intersects or contains another. The lower bounds are tight for special side-information configurations and for all problems with up to four absent receivers~\cite{OngConf3}. 

Other studies focused on devising optimal coding schemes for specific classes of PICOD($t$) problems. Liu and Tuninetti~\cite{liu2019tight} investigated a subset of PICOD($t$) problems known as complete-$S$ problems. In these problems, every size $s$ subset of $m$ messages where  $s\in S$  serves as the side information for a receiver. Optimal code length has been found for two classes of complete-$S$ problems: $S = \{s_\text{min}, \dotsc, s_\text{max}\}$ or $S = \{0, \dotsc, s_\text{min}-1 \} \cup \{s_\text{max}+1, \dotsc, m-1\}$, for any $s_\text{min} \leq s_\text{max}$. For these two classes, an optimal code either conveys messages in an uncoded fashion or is an MDS code (based on partial clique for index coding). The matching converse is mainly based on MAIS.

In this work, we consider a generalization of complete-$S$ PICOD($t$). Our generalization is termed $g$-group complete-$S$ PICOD($t$) problem where there are $m$ groups each of $g$ messages, and every receiver that has $s$ groups where $s\in S$ is present in the problem. In total, there are $\sum_{s\in S}\binom{m}{s}$ receivers. Our work focuses on the singleton $S=\{s\}$ case where $1\leq s\leq m$. We first present a multi-stage coding scheme that uses the ingredients of code design that are optimal in the ungrouped complete-$S$ setting. The scheme is shown to be optimal for $t>(g-1)(m-s)$. For $t\leq (g-1)(m-s)$, we establish that the scheme is optimal among all broadcast codes. Our results builds on and subsumes those of Liu and Tuninetti for the singleton S when we set $g=1$~\cite{liu2019tight}.

This paper is organized as follows. Section~\ref{sec:probform} formally presents the group complete PICOD problem, Section~\ref{sec:notation} presents the notation used in the problem and discusses the tools and relevant graph-theoretical results used in this work, Section~\ref{sec:results} presents our achievability, converse, and pertinent ancillary results. Most proofs are relegated to the supplementary appendix document.

\section{Problem Formulation}\label{sec:probform}
The Pliable Index Coding (PICOD($t$)) problem consists of the following components:
\begin{itemize}
    \item A server possesses $m$ i.i.d messages $X_1, X_2, \ldots, X_m$ each of which take values in $\mathbbm{F}^L$, where $\mathbbm{F}$ is a finite field, and $L$ is the number of symbols of $\mathbbm{F}$ in each message.
    \item $n$ receivers that each have a subset of $m$ messages (known as \emph{side information}), and request any additional $t$ messages not in their side information. Without loss of generality, we may assume that each receiver has at most $m-t$ messages in the side information. 
\end{itemize}
The goal of the PICOD($t$) problem is to design strategies for the server to compress the $m$ messages into an index that is broadcast over a noiseless channel to meet the receiver demands. A specific class of PICOD($t$) problems known as complete-$S$ PICOD($t$) problem was introduced in \cite{liu2019tight}. In this formulation, $S\subseteq \{1,2,\ldots, m-t\}$ defines the receivers and their side information as follows:
\begin{itemize}
\item For each $s\in S$, there are $\binom{m}{s}$ receivers each with a distinct subset of size $s$ of messages as side information. We denote the side information of node $r$ as $\mathcal{M}_r$.
\item In total, there are $\sum_{s\in S} \binom{m}{s}$ receivers. 
\end{itemize}

For example, if $m=3$, $t=1$ and $S=\{1,2\}$, then the complete-$S$ PICOD($t$) problem with $3$ messages consists of $\binom{3}{1}+\binom{3}{2}=6$ receivers $r_1,\ldots, r_6$ with side-information subsets $\mathcal{M}_{r_1}=\{X_1\}$, $\mathcal{M}_{r_2}=\{X_2\}$, $\mathcal{M}_{r_3}=\{X_3\}$, $\mathcal{M}_{r_4}=\{X_1,X_2\}$, $\mathcal{M}_{r_5}=\{X_1,X_3\}$, and $\mathcal{M}_{r_6}=\{X_2,X_3\}$, respectively. Since $t=1$, each receiver wants an additional message that it does not already know. 

This work focuses on a generalization of the complete-$S$ PICOD($t$) problem, referred to as \textit{$g$-Group Complete-$S$ PICOD($t$)} problem with $m$ groups defined as follows:
\begin{itemize}
\item There are $mg$ messages partitioned into $m$ groups of $g$ messages each. The groupings are given by $\mathcal{G}_i = \{X_{(i-1)g+1}, \ldots, X_{ig}\}$, $i=1,\ldots, m$.
\item $S\subseteq \{1,\ldots, \lfloor \frac{mg-t}{g} \rfloor\}$ defines the number and side information of the receivers in the problem instance. For every $s\in S$, there are $\binom{m}{s}$ receivers each with $s$ groups as side information. The set of all  $\sum_{s\in S} \binom{m}{s}$ receivers in the problem is denoted by $\mathfrak R$.

For example, if $m=3$, $g=2$ $S=\{1,2\}$, then the problem has $\binom{3}{1}+\binom{3}{2}$ receivers. The side information of the six receivers $r_1,\ldots, r_6$ are $\mathcal{M}_{r_1}=\mathcal{G}_1$, $\mathcal{M}_{r_2}=\mathcal{G}_2$, $\mathcal{M}_{r_3}=\mathcal{G}_3$, $\mathcal{M}_{r_4}=\mathcal{G}_1\cup\mathcal{G}_2$, $\mathcal{M}_{r_5}=\mathcal{G}_1\cup\mathcal{G}_3$, and $\mathcal{M}_{r_6}=\mathcal{G}_2\cup\mathcal{G}_3$ respectively. Observe that the receivers either have 2 messages (1 group) or 4 messages (2 groups) in their side information.
\item Each receiver requests $t$ extra messages that is not in its side information.
\end{itemize}
Note that if we choose $g=1$, our $g$-group complete-$S$  formulation reverts back to the complete-$S$ setup of \cite{liu2019tight}. Even though the receiver demands are \emph{pliable}, the server has to select which $t$ messages to communicate to which receiver. This is done through \emph{decoding choice} (function) $\mathcal{D}:\mathfrak{R}\rightarrow 2^{\llbracket mg\rrbracket}$ with $|\mathcal{D}(r)|=t$ for each $r\in\mathfrak{R}$. Throughout this paper, we use the notation that $\llbracket m \rrbracket$  as a shorthand for $\{1,2,\ldots, m\}$ for any natural number $m$, and for a set $X$, $2^X$ denotes the power set of $X$.

For the sake of analysis, it is sometimes useful to relax the condition that the decoding choice for each receiver is precisely $t$ messages. It is therefore convenient to define a \emph{generalized decoding choice} that assigns at least $t$ messages to each receiver. Thus, a generalized decoding choice is a function $\mathcal{D}:\mathfrak{R}\rightarrow 2^{\llbracket mg\rrbracket}$, where $|\mathcal{D}(r)|\geq t$ for each $r\in\mathfrak{R}$.

Given a PICOD problem, the server selects a decoding choice $\mathcal{D}$, and a \emph{pliable index code} $C$ that works for this choice, whose details are as follows:
\begin{itemize}
    \item $C$ comprises of an encoding function $\mathbbm{E}: \left(\mathbb{F}^L\right)^{mg} \rightarrow \mathbbm{F}^\ell$ that the server uses to jointly encode the $mg$ messages and convey $l$ symbols of $\mathbbm{F}$ to each receiver. The encoder transmits $\mathbbm{E}(X_1,\ldots, X_{mg})$ to all receivers noiselessly. The normalized size $\frac{\ell}{L}$ is termed the \emph{rate} $R_C$ of the code.
    \item Each receiver $r\in\mathfrak{R}$, $C$ comprises of a decoding function $\mathbbm{D}_r:\mathbbm{F}^\ell \times \left(\mathbbm{F}^L\right)^{\kappa_r}\rightarrow  \left(\mathbbm{F}^L\right)^t$, where $\kappa_r$ denotes the number of messages in the side information of receiver $r$. Each receiver $r\in\mathfrak{R}$,  uses the decoding function $\mathbbm{D}_r$ with the broadcast encoded message $\mathbbm{E}(X_1,\ldots, X_{mg})$ along with its side information to decode the $t$ messages in $\mathcal{D}(r)$, i.e., $\{X_j: j\in\mathcal{D}(r)\}$.
\end{itemize}

In this work, we focus solely on the canonical singleton setting of $g$-group complete-$S$ PICOD($t$) where $S=\{s\}$ for $s\in\llbracket \lfloor \frac{mg-t}{g} \rfloor\rrbracket$. In this problem setting, there are $\binom{m}{s}$ receivers each with $s$ groups of messages in its side information. The goal is to identify the optimal broadcast rate $R^*(m,s,g,t)$ that is the least among all pliable index codes, i.e., 
$$R^*(m,s,g,t) =\min_{\mathcal{D}}\left(\min_{C \textrm{ works for } \mathcal{D}} R_C\right).$$

\section{Required Graph-theoretic Notions and Tools} \label{sec:notation}

As in previous works,  non-achievability (lower bound) results are shown predominantly using graph-theoretic arguments with a heavy use of the maximum acyclic induced subgraph (MAIS) bound \cite{bar2011index}. We therefore first describe our notation for the graph representation. Given a $g$-group complete-$\{s\}$ PICOD($t$) problem with $m$ groups, and once a particular decoding choice $\mathcal{D}$ is selected, we can represent the resulting index coding problem as a directed graph $G_{\mathcal{D}}$ with $t\binom{m}{s}$ nodes using the following two rules: 
\begin{itemize}
\item[R1] A receiver $r$ with side information   $\mathcal{M}_r=\cup_{\ell=1}^s\mathcal{G}_{i_\ell}$ that requests messages $\mathcal{D}(r)=\{j_1,\ldots, j_t\}$ is represented by $t$ separate nodes labelled as $(r, j_1), (r, j_2), \ldots, (r, j_t)$.
\item[R2] A directed edge from node $(r,j)$ to $(r',j')$ exists in $G_{\mathcal{D}}$ iff $X_{j'}\in \mathcal{M}_{r}$, i.e., the node $(r,j)$ represents a receiver whose side information  contains $X_{j'}$.
\end{itemize}
In this digraph representation, a node $(r,j)$ is said to have \emph{receiver label} $r$, and (decoded) message label $j$. Conversely, given a receiver $r$ and a message index $j\in \mathcal{D}(r)$, we view $(r,j)$ as one of the $t$ distinct node representations of $r$. 

In the ungrouped ($g=1$) case, a receiver $r=\{i_1,\ldots,i_s\}$ that recovers $t$ messages has the side information of $\binom{t+s}{s}-1$ \textit{other} receivers, and can therefore emulate them. Here, by \textit{emulate}, we mean the receiver $r$ can possibly decode more messages by use of the decoding function of any other receiver whose side information it possesses after decoding its $t$ messages. This can be done iteratively until the receiver $r$ can emulate no more receivers. In the grouped case, it is not necessary that a node can emulate another upon decoding its $t$ messages. Instead, a receiver $r$ can emulate another receiver $r'$ upon decoding its $t$ messages iff the messages exclusive to the side information of $r'$ is decoded by $r$, i.e., $\mathcal{M}_{r'}\setminus \mathcal{M}_r\in \mathcal{D}(r)$. Therefore, a necessary and sufficient condition for $r$ to emulate another receiver is that $\mathcal{D}(r)$ contain at least another full group.

One of the key benefits of this digraph representation is that it allows us to focus on the sizes of acyclic induced subgraphs of $G_\mathcal{D}$ to bound from below the rate for the PICOD$(t)$ problem. In fact, the best such lower bound is given by the size of the \textit{maximum} acyclic induced subgraph with distinct message labels~\cite{bar2011index}. In this work, we denote $\texttt{MAIS}(G_{\mathcal{D}})$ to be any one \textit{maximum} acyclic induced subgraph of $G_\mathcal{D}$ with distinct message labels, and let $|\texttt{MAIS}(G_{\mathcal{D}})|$ to be its size. Given a PICOD problem, a decoding choice $\mathcal D$, and an MAIS $G$ in $G_{\mathcal{D}}$, we will find it handy to organize $G$ in layers based on the following rules:
\begin{itemize}
\item[R1$'$] Layer 1, denoted by $\mathcal{L}_1$, consists of nodes of $G$ that are sources (i.e., nodes with no incoming edges);
\item[R2$'$] For $i>1$, Layer $i$, denoted by $\mathcal{L}_i$, consists of all nodes whose in-neighbors consists of nodes in Layers $j$ for $j<i$ with at least one in-neighbor in Layer $i-1$; and
\item[R3$'$] The last layer consists only of sinks (but not necessarily all the sinks).
\end{itemize}
We say that a group $\mathcal G$ of messages is \emph{fully present} in a layer $\mathcal{L}_x$, if $\mathcal{G}\subseteq \{X_j: (r,j)\in \mathcal{L}_x\}$. We say a group $\mathcal G$ of messages is \emph{partially present} in a layer $\mathcal{L}_x$, if $0<|\mathcal{G}\cap \{X_j: (r,j)\in \mathcal{L}_x\}|<g$. Lastly, we say that a group $\mathcal G$ of messages is \emph{absent} in a layer $\mathcal{L}_x$, if $\mathcal{G}\cap \{X_j: (r,j)\in \mathcal{L}_x\}=\emptyset$. Conversely, a layer $\mathcal{L}_x$ is \emph{perfectly covered} by groups if there exists a collection of groups $\mathcal{G}_{i_1},\ldots, \mathcal{G}_{i_k}$ whose $kg$ message labels together form the message labels of the nodes in the layer, i.e., $\mathcal{G}_{i_1}\cup\ldots\cup \mathcal{G}_{i_k}=\{X_j: (r,j)\in \mathcal{L}_x\}$. We extend the above notions to say a group $\mathcal G$ of messages is \emph{partially present} or \emph{fully present} in a MAIS $G$ of $G_{\mathcal{D}}$ if it is partially present or fully present in any one layer. Note that from Property P5 below, a group cannot be partially present in more than one layer. Similarly, we say that a group $\mathcal G$ of messages is absent from a MAIS $G$ of $G_{\mathcal{D}}$ if it is absent from every layer of $G$.

\begin{lemma}\label{lem:GraphProperties} The following properties are true of any MAIS $G$ of any digraph representation $G_{\mathcal{D}}$.
\begin{itemize}
    \item[P1] An edge starting from a node of Layer $i$ can only end in a node in a layer of larger index.
    \item[P2] If $j$ is the message label of a node in Layer $i$, and $r$ is receiver label of a node in Layer $k$ for $k\geq i$, then the side information of $r$ does not contain $X_j$.
    \item[P3] Two nodes that have the same receiver label have the same out-neighborhood.
    \item[P4] Two nodes whose message labels are from the same group have the same in-neighborhood. 
    \item[P5] A group can be partially present in only one layer. 
    \item[P6] The set of message labels in an MAIS and the subset of messages in the side information of the sinks are disjoint.
    \end{itemize}
\end{lemma}
\begin{IEEEproof}
The proof of the properties are as follows:
\vspace{1mm} 

\noindent{}P1:  An edge from a node in Layer $k$ to a node in Layer $i$ for $k>i$ would violate Rule R2$'$ of the construction of layers. 
\vspace{1mm} 

\noindent{}P2: Suppose $X_j\in\mathcal{M}_r$. Then there must be an edge from the node with receiver label $r$ in Layer $k$ to the node in Layer $i$ with message label $j$. This then violates P1 since $k\geq i$.\vspace{1mm} 

\noindent{}P3: The condition for an edge to start in a node $(r,j)$ and end in a node $(r',j')$ depends only on the receiver label $r$ and the message label $j'$. Hence any two nodes that have the same receiver label must have the same out-neighborhood. 
\vspace{1mm} 

\noindent{}P4: From Rule R2 of digraph construction, it is clear that the in-neighborhood of a node $(r,j)$ is uniquely determined by the group where $X_j$ lies. Thus, two nodes with message labels from the same group have the same in-neighborhood.
\vspace{1mm} 

\noindent{}P5: Suppose that there are $(r,j)$ and $(r',j')$ in $G$ such that: (a) $X_{j}, X_{j'}$ belong to the same group; and (b) the nodes  $(r,j)$ and $(r',j')$ lie in layers $k$ and $k'$, respectively, with $k\leq k'$. From P4, we see that the in-neighborhood of $(r,j)$ and $(r',j')$ are the same. Hence, they cannot lie in different layers. 
\vspace{1mm} 

\noindent{}P6: If they are not disjoint, a message index is present both as message label and in the side-information of a sink node. By Rule R2, this would mean that the sink has an outgoing edge, which is a contradiction.  
\end{IEEEproof} 

 \begin{corollary}
 From Property P1 of Lemma~\ref{lem:GraphProperties}, it follows that there are no edges between nodes of the same layer.
 \end{corollary}
 \begin{corollary}
From Property P2 of Lemma~\ref{lem:GraphProperties}, it follows that if $j$ is a message label of a source node in an MAIS $G$ of $G_{\mathcal{D}}$, and $r$ is a receiver label of any node in $G$, then $X_j\notin \mathcal{M}_r$; and if $j$ is a message label of a node in $G$, and $r$ is a receiver label of a sink node in $G$, then again $X_j\notin \mathcal{M}_r$.
\end{corollary}

The last fact that we will need for our non-achievability results characterizes the size of an MAIS for generalized decoding choice obtained by appending to decoding choices messages that a receiver decodes by emulation. 
\begin{lemma}\label{lem:AppendDecChoice}
Let $\mathcal{D}$ be a decoding choice for a $g$-group complete-$\{s\}$ PICOD($t$)
problem with $m$ groups. Suppose a receiver $r$ uses $\mathbb D_r$ to decode its $t$ messages, and is then able to emulate $r_1$ to decode a new message $X_j$, where $j\in \mathcal{D}(r_1)$. Let $\mathcal{D}_{\textrm{app}}$ be a generalized decoding choice such that:
\begin{align*}
\mathcal{D}_{\textrm{app}}(x) = \left\{\begin{array}{cc} \mathcal{D}(x) & x\neq r\\ \mathcal{D}(x)\cup\{j\} & x=r \end{array}\right.. 
\end{align*}
Then, $|\texttt{MAIS}(G_{\mathcal{D}_{\textrm{app}}})| = |\texttt{MAIS}(G_{\mathcal{D}})|. $ 
\end{lemma}
\begin{IEEEproof}
Proof is given in the supplementary document.
\end{IEEEproof}
\begin{remark}
Given a decoding choice $\mathcal{D}$, we can sequentially append for each receiver every additional messages it will decode by emulating other receivers. We repeat this until we can add no new messages to any receiver. Let $\overline{\mathcal D}$ be the generalized decoding choice upon termination of this process. By repeated application of the Lemma~\ref{lem:AppendDecChoice}, we can show that
$$|\texttt{MAIS}(G_{\overline{\mathcal{D}}})|=|\texttt{MAIS}(G_{\mathcal{D}})|.$$
\end{remark}
\begin{corollary}\label{cor:appendmessages}
Given a decoding choice $\mathcal{D}$, let $\mathcal{T} =\{(r_k,j_k)\}_{k=1}^{\ell}$ be an enumeration of the additional messages decoded by receivers by successive emulation. Suppose we append $\mathcal{D}$ with a subset $\mathcal{S}\subseteq \mathcal{T}$ of extra messages decoded by receivers to obtain a generalized decoding choice $\mathcal{D}^*$. Then, by a similar argument, we can show that
$$|\texttt{MAIS}(G_{\overline{\mathcal{D}}})|= |\texttt{MAIS}(G_{{\mathcal{D}^*}})|=|\texttt{MAIS}(G_{\mathcal{D}})|.$$
\end{corollary}

\section{New Results}\label{sec:results}
We begin this section with the achievability result, which is an extension of the one proposed in \cite{liu2019tight}.
\begin{theorem}\label{thm:achievable}
$R^*(m,s,g,t) \leq \min\{s+t, \lceil\frac{t}{(m-s)}\rceil (m-s)\}$.
\end{theorem}
\begin{IEEEproof}
The code construction for a $g$-group complete-$\{s\}$ PICOD$(t)$ problem with $m$ groups is as follows:
\begin{itemize}
\item For each $i=1,2,\ldots, \lfloor\frac{t}{m-s}\rfloor$, the sender conveys $\{ X_i, X_{g+i}, \ldots, X_{(m - 1)g + i}\}$ (one new message from each of the $m$ groups) via an MDS code of size $m-s$. Each receiver already possess $s$ of these $m$ encoded messages, and will see the $s$ messages to decode the remaining $m-s$ messages. At the end of these rounds, every receiver will have decoded a total of $\lfloor\frac{t}{m-s}\rfloor (m-s)$ messages with $\lfloor\frac{t}{m-s}\rfloor$ messages from each group.
\item If $t'=t-\lfloor\frac{t}{m-s}\rfloor (m-s)>0$, then sender does one of the following:
\begin{itemize}
\item If $s+t'<m-s$, the sender transmits messages $X_{j},X_{g+j},\ldots, X_{(s+t'-1)g+j}$ uncoded, where $j= \lfloor\frac{t}{m-s}\rfloor+1$.
\item If $s+t'\geq m-s$, the sender encodes $X_{j},X_{g+j},\ldots, X_{(m-1)g+j}$ (one new message from each of the $m$ groups) via an MDS code of size $m-s$. Each receiver then receives $m-s$ new messages, which may be more than the remaining $t'$ they need. 
\end{itemize}
\end{itemize}
\end{IEEEproof}
The following figure presents the plot of the achievable rate offered by Theorem~\ref{thm:achievable}. Additionally, the applicable matching non-achievability results for various ranges of $t$ are also indicated therein. Note from the proof of achievability that in the last round, an MDS code will be used if and only if $s\geq m-s$, in which case the achievable rate is given by $\lceil \frac{t}{m-s}\rceil (m-s)$; alternately, the last round will comprise of uncoded messages iff $s<m-s$, which introduces the unity slope portions. In both cases, the achievable rate region is non-convex due to the absence of a time-sharing argument that is usually used to convexify rate regions. 
\begin{figure}[!ht]
  \centering 
  \includegraphics[width=3.3in]{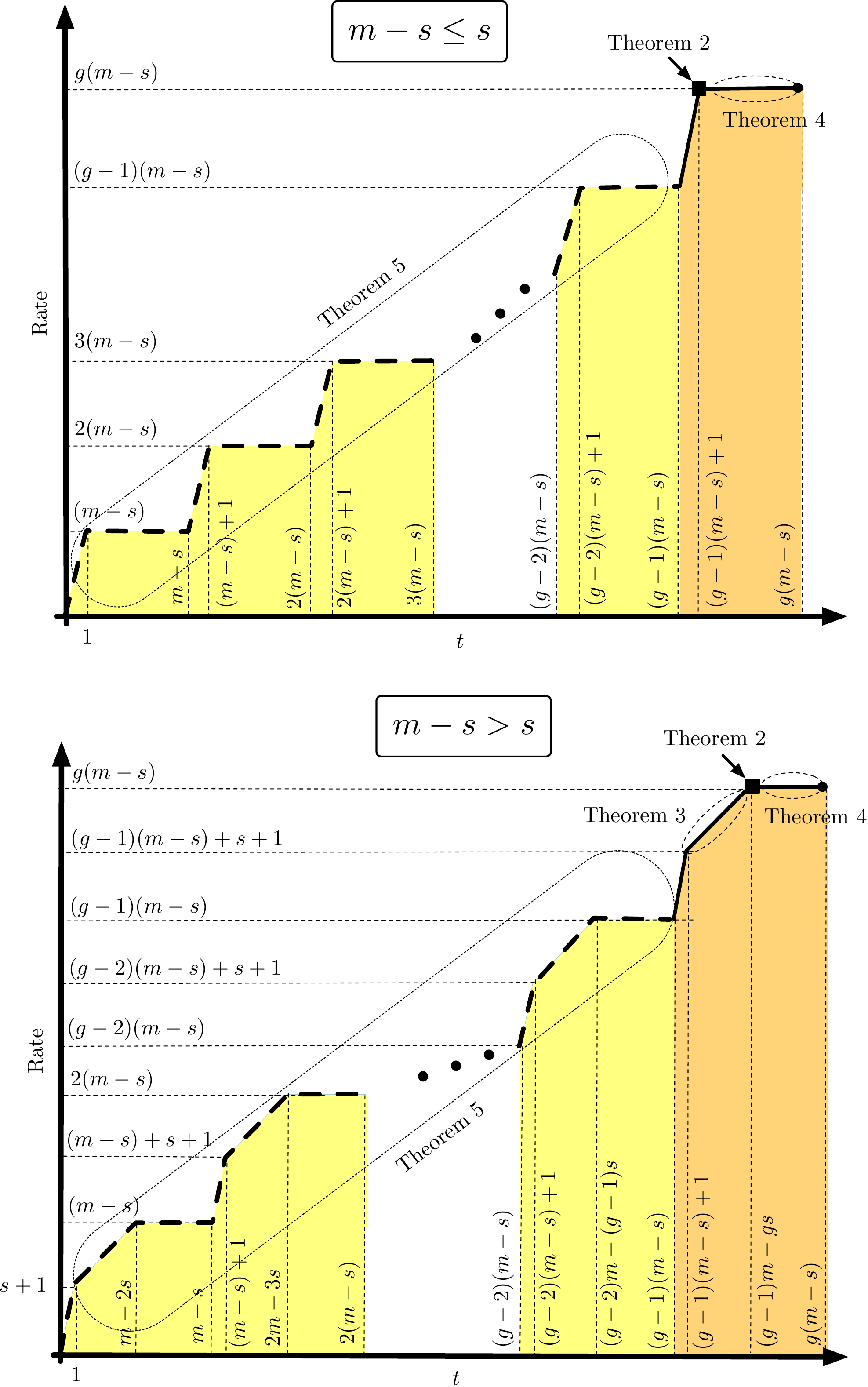}
\caption{The achievable rate given in Theorem~\ref{thm:achievable}}
\end{figure}
We now present three ancillary results that are needed to derive our non-achievability results. The first two connect the size of MAIS of $G_{\mathcal {D}}$ to the number of messages decoded by a node in $G_{\mathcal {D}}$. The third result argues that under certain conditions if there is a decoding choice whose size of a MAIS is $s+t-i$ for $i>0$, then there must exist another decoding choice whose size of MAIS is $s+t-i+1$. Our non-achievability results prove the non-existence of a decoding choice whose MAIS size is $s+t-1$, and therefore, using the third result, we are guaranteed that the smallest size of any MAIS of any decoding choice is $s+t$.

\begin{lemma}\label{lem:decodetomais}
If for a decoding choice $\mathcal{D}$, there exists a receiver $r$ that decodes $s+t$ messages by sequentially emulating other receivers, then $|\texttt{MAIS}(G_{\mathcal{D}})|\geq s+t$.
\end{lemma}
\begin{IEEEproof}
Let $\mathcal{D}(r)=\{j_1,\ldots, j_t\}$. Then $r$ uses the decoding function  $\mathbbm{D}_r$ of the receiver $r$ to decode $X_{j_1},\ldots, X_{j_t}$. Since, $r$ decodes $s+t$ messages by emulating other receivers, these $t$ messages must contain at least one full group allowing it to emulate another receiver. This process repeats until $r$ decodes $s$ additional messages. Let $X_{j_1},\ldots, X_{j_{s+t}}$ be an enumeration of the $s+t$ distinct messages $r$ decodes sequentially. For $i=1,\ldots, s+t$, let $r_{k_i}$ be the receiver that $r$ emulates to decode $X_{j_i}$ using $\mathbbm{D}_{r_i}$. Then, the following statements hold.
\begin{itemize}
\item[F1] For any $i\in\llbracket s+t\rrbracket $,  $(r_{k_i}, j_i)$ is a node in $G_\mathcal{D}$.
\item[F2] For any $i\in\llbracket s+t\rrbracket$, $\mathcal{M}_{r_{k_i}}\subseteq \mathcal{M}_r \cup \{X_{j_1},\ldots, X_{j_{i-1}}\}$.
\item[F3] For any $i\in\llbracket s+t\rrbracket$, $X_{j_i}\notin \mathcal{M}_r$. 
\item[F4] For any $0<\ell<\ell'\leq s+t$, $X_{\ell'}\notin \mathcal{M}_{r_{k_\ell}}$. Otherwise, $X_{\ell'}$ must either be decoded earlier in the enumeration of messages decoded by $r$, or is in the side information of $r$. The former inference violates the fact that the $s+t$ messages are distinct, the latter violates the fact that $X_{\ell'}$ is innovative to $r$.
\end{itemize}
Consider the subgraph induced by $\{v_i=(r_{k_i},j_i): i=1,\ldots, s+t\}$ in $G_{\mathcal{D}}$. From F4 above, we see that this induced subgraph has edges from $v_i$ to $v_j$ only if $i<j$. Consequently, this subgraph of $s+t$ nodes is acyclic with distinct message labels, and therefore, the claim follows. 
\end{IEEEproof}

\begin{lemma}\label{lem:maistodecode}
Let $m,s,g,t$ be such that $t>(g-1)(m-s)$ and $s+t=g(m-s)$. If for a decoding choice $\mathcal{D}$, $|\texttt{MAIS}(G_{\mathcal{D}})|= s+t$, then there exists a receiver $r$ that decodes $s+t$ messages.
\end{lemma}
\begin{IEEEproof}
Let $v_k=(r_{i_k},j_{k})$, $k=1,\ldots, s+t$, be the labels of an MAIS $G_0$ of $G_{\mathcal D}$. Since this subgraph is acyclic, we may assume that $v_1,v_2,\ldots, v_{s+t}$ is in a topologically sorted order with sources at the beginning and the sinks at the end. Now, consider the sink $v_{s+t}=(r_{i_{s+t}},j_{{s+t}})$. Observe that the $mg$ messages in the problem are partitioned into two: the $gs$ messages that are in the side information of receiver $r_{i_{s+t}}$, and the $(s+t)=g(m-s)$ message labels of the nodes in $G_0$. Therefore, it follows that all sinks of $G_0$ must have the same receiver label $r_{i_{s+t}}$. Now, since $r_{i_{s+t}}$ decodes $t>(g-1)(m-s)$ messages, it must decode at least one full group. Now, consider the acyclic subgraph $G_1$ obtained by deleting all nodes in $G_0$ whose message label is decoded by $r_{i_{s+t}}$ by use of $\mathbbm{D}_r$. Consider the sinks of $G_1$. As before, the message labels in $G_1$ is precisely $\mathcal{M}_{r_{i_{s+t}}}^c \setminus \mathcal{D}(r_{i_{s+t}})$ (due to Property P6 of Lemma~\ref{lem:GraphProperties}). Thus, the sinks of $G_1$ whose side information is a subset of the complement of the message labels in $G_1$, can each be emulated by $r_{i_{s+t}}$ to uncover more messages. 

This process of emulating and recovering more messages can be followed iteratively until the receiver labels of all nodes in $G_0$ are emulated by $r_{i_{s+t}}$. Thus, at each iteration $k>0$, we can define a new acyclic subgraph $G_k$ by deleting all nodes in $G_{k-1}$ whose message label is decoded by the sink nodes of $G_{k-1}$ by use of their decoding functions. Since $r_{i_{s+t}}$ can emulate the sink nodes of $G_{k-1}$, $r_{i_{s+t}}$ uncovers all messages decoded by the sink nodes of $G_{k-1}$. Thus, we see that the receiver $r_{i_{s+t}}$ will eventually decode the $s+t$ messages corresponding to the message labels of the nodes in $G_0$.
\end{IEEEproof}

\begin{lemma}\label{lem:IncrementByOne}
Let $m,s,g,t$ be such that $t>(g-1)(m-s)$ and $s+t\leq g(m-s)$. Then, 
if there exists a decoding choice $\mathcal{D}$ with $|\texttt{MAIS}(G_\mathcal{D})| = s + t - i $ for $i > 0$ then there also exists $ \mathcal{D}'$ with $|\texttt{MAIS}(G_{\mathcal{D}'})| = s + t - i + 1$.
\end{lemma}
\begin{IEEEproof}
The proof proceeds in three cases depending on the number of sinks $k$ in $G_{\mathcal{D}}$. The case that $k>t$ is shown to lead to contradiction, whereas in the other two cases, a specific node $(r,j^*)$ in $G_{\mathcal{D}}$ and a message label $\hat j\notin \mathcal{D}(r)$ are identified. The decoding choice $\mathcal{D}$ is then altered to obtain $\hat{\mathcal{D}}$ by replacing $j^*$ in $\mathcal{D}(r)$ with $\hat j$. It is then shown that $|\texttt{MAIS}(G_{\hat{\mathcal{D}}})| = |\texttt{MAIS}(G_{\mathcal D})|+1$. The proof is detailed in the supplementary appendix document. 
\end{IEEEproof}
\begin{corollary}
\label{cor:sabsentgroups}
Let $m,s,g,t$ be such that $t>(g-1)(m-s)$ and $s+t\leq g(m-s)$. From  Lemma~\ref{lem:IncrementByOne} Case 1, it is clear that precisely $s$ groups are absent from an $\texttt{MAIS}(G_{\mathcal{D}})$ for any $\mathcal{D}$.
\end{corollary}

We are now ready to present our non-achievability results.
\begin{theorem} \label{thm:lastcriticalpt}
Let $m,s,g,t$ be such that $t>(g-1)(m-s)$ and $s+t= g(m-s)$. Then, $R^*(m,s,g,t)=s+t$.
\end{theorem}
\begin{IEEEproof}
 The proof starts with a decoding choice $\mathcal{D}$ for which the size of MAIS is $s+t-1$. It identifies a special receiver $
 \hat r$ that can decode an additional message by emulating other receivers. Constructing a generalized decoding choice by appending to $\mathcal{D}(\hat r)$ this additional message allows to then establish that an MAIS corresponding to decoding choice $\mathcal D$ has to have a size at least $s+t$. By invoking Lemma~\ref{lem:IncrementByOne}, we establish  that $s+t$ is a lower bound to the size of an MAIS corresponding to any decoding choice. Complete proof can be found in the supplementary appendix document. 
\end{IEEEproof}

\begin{theorem}\label{thm:belowlastcriticalpt}
    Let $m,s,g,t$ be such that $t>(g-1)(m-s)$ and $s+t< g(m-s)$. Then, $R^*(m,s,g,t)=s+t$.
\end{theorem}
\begin{IEEEproof}
 The proof follows closely that of Theorem~\ref{thm:lastcriticalpt}, and can be found in the supplementary appendix document. 
\end{IEEEproof}
\begin{theorem}\label{thm:abovelastcriticalpt}
Let $m,s,g,t$ be such that $s+t>g(m-s)$ and $g-1<\frac{t}{m-s}\leq g$. Then, $R^*(m,s,g,t)=g(m-s)$.
\end{theorem}
\begin{IEEEproof}
As in the proof of Proposition~7 of \cite{liu2019tight}, this proof selects a sub-problem that is a $g$-group complete-$\{s'\}$ problem over $m'$ groups for suitable $m' $ and $g'$ for which Theorem~\ref{thm:lastcriticalpt} is applicable. Complete proof can be found in the supplementary appendix document. 
\end{IEEEproof}

\begin{theorem}\label{thm:restrictedconverse}
Let $\mathcal{C}_{m,s,g,t}$ denote the family of \emph{broadcast} codes for the $g$-group complete-$\{s\}$ PICOD($t)$ problem with $m$ groups  where if a message is conveyed to one receiver, every receiver that does not have that message in its side information also decodes that message. The optimal rate among codes from $\mathcal{C}_{m,s,g,t}$ for any $(f-1)(m-s)<t<f(m-s)$ for $f\in\llbracket g\rrbracket$ is given by $R'(m,s,g,t) = \min\{s+t,f(m-s)\}.$
\end{theorem}
\begin{IEEEproof}
The proof establishes that among broadcast codes, there exists an optimal one where the sender \emph{ignores} $g-f$ messages and communicates at most $f$ messages per group. This then allows us to reduce the problem to a $f$-group complete-$\{s\}$ PICOD($t$) problem whose rate can be deduced from Theorems~\ref{thm:belowlastcriticalpt} and \ref{thm:abovelastcriticalpt}. Complete proof can be found in the supplementary appendix document. 
\end{IEEEproof}



\newpage

\newpage 
$\,$
\newpage
\appendices
\section*{\Large Appendix}
\section*{Proof of Lemma~\ref{lem:AppendDecChoice}}
 Note that, by construction, $G_{\mathcal{D}}$ is a subgraph of $G_{\mathcal{D}_{\textrm{app}}}$ since the latter contains all nodes and edges of $G_{\mathcal{D}}$ and additionally one extra node $(r,j)$ and its incident incoming and outgoing edges. Furthermore, $G_{\mathcal{D}}$ is an induced subgraph of $G_{\mathcal{D}_{\textrm{app}}}$ since $G_{\mathcal{D}_{\textrm{app}}}$ does not have any additional edges among nodes already present in $G_{\mathcal{D}}$. Therefore, it follows that any (acyclic) induced subgraph of $G_{\mathcal{D}}$ is also an (acyclic) induced subgraph of $G_{\mathcal{D}_{\textrm{app}}}$. However, $G_{\mathcal{D}_{\textrm{app}}}$ could potentially have acyclic induced subgraphs that involve the node $(r,j)$. Thus it follows that  $|\texttt{MAIS}(G_{\mathcal{D}_{\textrm{app}}})| \geq |\texttt{MAIS}(G_{\mathcal{D}})|$. Next, to show $|\texttt{MAIS}(G_{\mathcal{D}_{\textrm{app}}})| = |\texttt{MAIS}(G_{\mathcal{D}})|$, it suffices to establish $|\texttt{MAIS}(G_{\mathcal{D}_{\textrm{app}}})| \leq |\texttt{MAIS}(G_{\mathcal{D}})|$.

 The only extra components $G_{\mathcal{D}_{\textrm{app}}}$ contains is the node $(r,j)$ and its incident edges. Now, let $G_{\textrm{app}}$ be an MAIS for $G_{\mathcal{D}_{\textrm{app}}}$. If $G_{\textrm{app}}$ does not contain the node $(r,j)$ it is also an MAIS for $G_{\mathcal{D}}$, and the claim follows. We may therefore assume that $(r,j)$ is present in $G_{\textrm{app}}$. We now proceed in cases.
    \vspace{1mm}
    
    \noindent{} Case A: \textit{\textrm{(r,j)} is a source (i.e., in Layer $\mathcal{L}_1$) of $G_{\textrm{app}}$}. In this case, consider another induced subgraph $G'_{\textrm{app}}$ constructed from $G_{\textrm{app}}$ by deleting $(r,j)$ and its incident edges from $G_{\textrm{app}}$, and adding $(r_1,j)$ and required incident edges (to make the subgraph induced). Since originally, $(r,j)$ was a source in $G_{\textrm{app}}$, the receiver label of every other node in $G_{\textrm{app}}$ did not contain $X_j$ in its side information. Therefore, $(r_1,j)$ has to be a source node in $G'_{\textrm{app}}$. This further means that $G_{\textrm{app}}'$ is acyclic, since any newly introduced loop in $G_{\textrm{app}}'$ must traverse $(r_1,j)$, but that is impossible since $(r_1,j)$ is a source node in $G'_{\textrm{app}}$. Thus, $G'_{\textrm{app}}$ is an acyclic induced subgraph of $G_{\mathcal{D}_{\textrm{app}}}$ whose size is the same as $G_{\textrm{app}}$. Thus, $G'_{\textrm{app}}$ is also an MAIS of $G_{\mathcal{D}_{\textrm{app}}}$. However, since $G'_{\textrm{app}}$ does not contain $(r,j)$, it is also an MAIS of $G_{\mathcal{D}}$, and therefore, $|\texttt{MAIS}(G_{\mathcal{D}_{\textrm{app}}})| = |\texttt{MAIS}(G_{\mathcal{D}})|$.
    \vspace{1mm}
        
    \noindent{}Case B: \textit{\textrm{(r,j)} is not a source (i.e., in Layer $\mathcal{L}_\ell$ for $\ell>1$) of $G_{\textrm{app}}$}. In this case we construct an alternate MAIS for $G_{\mathcal{D}_{\textrm{app}}}$ as follows. As before, we start with $G_{\textrm{app}}$ and delete $(r,j)$ and its incident edges, and add $(r_1,j)$ and required incident edges (to make the subgraph induced). Let the resultant graph be $G'_{\textrm{app}}$. Now, Since $(r,j)$ originally was in layer $\mathcal L_\ell$, it follows that all its in-neighbors are in layers $\mathcal L_k$, $k<\ell$. Therefore, all incoming edges into $(r_1,j)$ after replacement will also be from earlier layers $\mathcal L_j$, $j<\ell$. However, it is possible that after replacement $G'_{\textrm{app}}$ contains edges from $(r_1,j)$ to some $k$ nodes in the previous layers. Let $(r_1', j_1), (r_2', j_2), \ldots, (r_k', j_k)$ denote these out-neighbors of $(r_1,j)$ in layers $\mathcal L_k$, $k<\ell$. These \emph{back} edges have the potential to induce cycles. Note that $j_1,\ldots, j_k$ are messages that $r_1$ possesses as side information that $r$ does not have in its side information. However, since $r$ emulates $r_1$ upon decoding its $t$ messages, $d_1,\ldots, d_k\in \mathcal{D}(r)$. Hence, we can now delete the $k$ nodes $(r_i',j_i)$ and their incident edges, and add $(r,j_i)$, $i\in\llbracket k \rrbracket$ and their incident edges instead. Let $G{''}_{\textrm{app}}$ denote the resultant graph.

    Thus far, the modifications from $G_{\textrm{app}}$ to $G_{\textrm{app}}^{''}$ have preserved the size and the induced nature of the subgraphs. The last step is to establish its acyclicity. This follows since any cycle in $G{''}_{\textrm{app}}$ must involve $(r_1,j)$, back edges and at least one of $(r,j_i)$'s. But note that the out-neighbor of $(r,j_i)$'s are nodes whose message labels are present in the side information of $r$, and such nodes are in layers $\mathcal{L}_{k}$, $k>\ell$, and since these layers are untouched, there are no back edges from nodes in these layers back to $(r_1,j)$. Therefore, $G_{\textrm{app}}^{''}$ is an MAIS of $G_{\mathcal{D}_{\textrm{app}}}$, and since the nodes of $G_{\textrm{app}}^{''}$ are all present in $G_{\mathcal{D}}$, it follows that $G_{\textrm{app}}^{''}$ is also an MAIS of $G_{\mathcal{D}}$, and therefore, $|\texttt{MAIS}(G_{\mathcal{D}_{\textrm{app}}})| = |\texttt{MAIS}(G_{\mathcal{D}})|$.

\section*{Proof of Lemma~\ref{lem:IncrementByOne}}
We have three cases depending on $G=\texttt{MAIS}(G_\mathcal{D})$.
\vspace{1mm}

\noindent{}Case 1: \textit{There are $k>t$ sinks in $G$.} In this case, since there are only $t$ node representations per receiver, there has to be at least two distinct sink node receiver labels. It is therefore the case that there are at least $s+1$ groups, say $\mathcal{G}_{i_1},\ldots,\mathcal{G}_{i_{s+1}}$ whose messages are present in the side information of the receivers represented by the sink nodes. Note that there are $t\binom{s+1}{s}= t(s+1)>s+t-i$ node representations of receivers whose side information is a subset of $\cup_{k=1}^{s+1}\mathcal{G}_{i_k}$. Let $(r,j)$ be one such node representation that is absent from $\texttt{MAIS}(G_\mathcal{D})$. Consider the subgraph $G'$ induced by vertices already in $G$ along with $(r,j)$. Note that $G'$ is also acyclic, since any cycle that is newly introduced has to traverse $(r,j)$; however, the out-degree of $(r,j)$ in $G'$ is zero, since neither is $X_j\in \mathcal{M}_r$ nor is a message label of a node in $G$ a message in $\mathcal{M}_r$. Hence, $G'$ is a larger acyclic induced subgraph subsuming $G$, which contradicts the maximality of $G$. Hence, this case cannot arise. 
\vspace{1mm}
    
\noindent{}Case 2: \textit{There are $k<t$ sinks in $G$.} If there are two distinct receiver labels among sink nodes, the union of the side information of sink nodes contains at least $s+1$ groups, which leads to a contradiction just as in Case 1. Thus, all sink nodes must have the same receiver label. Let $(r,j_1), \ldots, (r,j_k)$ denote the sink nodes. Let $j^*\in \mathcal{D}(r) \setminus\{j_1,\ldots, j_k\}$. Pick $\hat j\notin \mathcal{M}_r$ such that $\hat j$ does not appear as a message label in $G$. There are at least $i>0$ choices for $\hat j$, since the number of messages in $\mathcal{M}_r$ is $gs$, and the number of message labels is $s+t-i\leq g(m-s)-i$. Define a new decoding choice $\hat{\mathcal{D}}$ by:
    $$\hat{\mathcal{D}}(x) = \left\{\begin{array}{cc} \mathcal{D}(x) & x\neq r\\ \left(\mathcal{D}(r)\setminus\{j^*\}\right)\cup\{\hat j\} & x=r \end{array}\right..$$
 The decoding choice $\hat{\mathcal{D}}$ is the same for all receivers, but for $r$; and for $r$, only one message $j^*$ that possibly appears as a node label in $G$ is replaced by $\hat j$ that does not. Note that $G$ is also an acyclic induced subgraph of $G_{\hat{\mathcal{D}}}$; however, the subgraph $\hat G$ induced by vertices already in $G$ along with $(r,\hat j)$ is also an acyclic induced subgraph of $G_{\hat{\mathcal{D}}}$ since $(r,\hat j)$ is a sink node in $G_{\hat{\mathcal{D}}}$. In fact, $\hat G$ is maximal since changing only one message label for the decoding choice of only one receiver can alter the size of MAIS by at most one. The claim then follows since $|\hat G| =|\texttt{MAIS}(G_{\hat{\mathcal{D}}})|=s+t-i+1$.
 \vspace{1mm}
 
 \noindent{}Case 3: \textit{There are precisely $t$ sinks in $G$.} In this case, we look at $G$ through the layering given in Section~\ref{sec:notation}. We search $G$ to find the layer with the smallest index $\ell$ that contains a partially covered group, say $\mathcal{G}^*$. Now, four cases arise:
\begin{itemize}
\item[3a] \textit{$G$ has no partially covered group}. In this case, the message labels of nodes in $G$ is precisely a union of message labels of groups $\mathcal{G}_{i_1},\ldots, \mathcal{G}_{i_k}$, i.e., $\{j: (r,j)\in G\} = \{j: X_j \in \mathcal{G}_{i_1}\cup \ldots\cup \mathcal{G}_{i_k}\}$. Then, since $kg=|G|=s+t-i\leq g(m-s)-i$, it follows that $k<m-s$. Then, the $m-k>s$ groups are absent from the message labels of nodes in $G$. As in Case 1, there then are $t\binom{s+1}{s}$ node representations of receivers whose side information is a subset of these $s+1$ receivers. Since $t\binom{s+1}{s}>s+t-i$, one can find a node representation that is absent in $G$, and add to $G$ the representation and necessary incident edges to derive a larger acyclic induced subgraph leading then to a contradiction. So, this case does not arise. 

\item[3b] \textit{Partially covered group $\mathcal G^*$ is in layer $\mathcal{L}_{\ell}$, and $\mathcal{L}_{\ell}$ has a non-sink node with a message label in $\mathcal{G}^*$.} In this case, pick a non-sink node $(r,j^*)$ in this layer such that $X_{j^*}\in \mathcal{G}^*$. From P3 of Lemma~\ref{lem:GraphProperties}, it follows that every node with receiver label $r$ is also a non-sink node. There cannot be $t$ nodes in $G$ with receiver label $r$. Otherwise, $G$ has at least $t$ non-sink nodes and $t$ sink nodes, and $s+t-i=|G| \geq 2t > 2(g-1)(m-s) \geq g(m-s)$, which violates the hypothesis $s+t\leq g(m-s)$. Here, we have used the fact that $g\geq 2$ if a group is present partially. Let $(r,\tilde j)$ be a node in $G_{\mathcal{D}}$ that is not in $G$. 

Now, pick $\hat j$ such that $X_{\hat j}\in \mathcal{G}^*$ and $\hat j$ is not a message label in layer $\mathcal{L}_\ell$. Clearly, $\hat j\notin \mathcal{D}(r)$. Since, if it were, the induced subgraph with nodes of $G$ and $(r,\hat j)$ will be acyclic. This is because if there is a cycle involving $(r,\hat j)$, there will also be a cycle involving $(r, j^*)$ as their in- and out-neighborhoods are identical (due to Properties P3 and P4 of Lemma~\ref{lem:GraphProperties}), which contradicts the acyclicity of $G$. Now, define a new decoding choice $\hat{\mathcal{D}}$ as follows:
    $$\hat{\mathcal{D}}(x) = \left\{\begin{array}{cc} \mathcal{D}(x) & x\neq r\\ 
    \left(\mathcal{D}(r)\setminus\{\tilde j\}\right)\cup\{\hat j\} & x=r\end{array}\right..$$

Note that $\mathcal D$ and $\hat{\mathcal{D}}$ differ at most in one decoded message for precisely one receiver alone. Consider $\hat{G}$ to be the induced subgraph with nodes of $G$ and $(r,\hat j)$. By the above argument, $\hat{G}$ is also an acyclic induced subgraph of $G_{\hat{\mathcal{D}}}$; $\hat{G}$ is maximal since changing only one message label for the decoding choice of only one receiver can alter the size of MAIS by at most one. The claim then follows since $|\hat{G}| =|\texttt{MAIS}(G_{\hat{\mathcal{D}}})|=s+t-i+1$.

\item[3c] \textit{Partially covered group $\mathcal G^*$ is in layer $\mathcal{L}_{\ell}$, $\ell> 1$, and all nodes with message labels in $\mathcal{G}^*$ are sink nodes.}

Pick a node in layer $\mathcal{L}_{\ell}$, whose message label is in $\mathcal G^*$, and select one of its in-neighbor, say $(r^*,j^*)$, from layer $\mathcal{L}_{\ell-1}$. By an argument similar to Case 3b, we can infer that: (a) all nodes with receiver label $r^*$ are non-sink nodes, and (b) there cannot be $t$ node representations of the receiver $r^*$. Thus, there must be a representation $(r^*,\tilde j)$ in $G_{\mathcal D}$ that is not in $G$. Let $\hat j$ be a message label such that $X_{\hat j}\in\mathcal{G}^*$, and $\hat j$ is not a message label in $G$. Note that $\hat j \notin \mathcal{D}(r^*)$ using the same argument in Case 3b. Now, define a new decoding choice $\hat{\mathcal{D}}$ as follows:
    $$\hat{\mathcal{D}}(x) = \left\{\begin{array}{cc} \mathcal{D}(x) & x\neq r^*\\ 
    \left(\mathcal{D}(r^*)\setminus\{\tilde j\}\right)\cup\{\hat j\} & x=r^*\end{array}\right..$$

Note that $\mathcal D$ and $\hat{\mathcal{D}}$ differ at most in one decoded message for precisely one receiver alone. Consider the induced subgraph $\hat{G}$ with nodes of $G$ and $(r^*,\hat j)$. $\hat{G}$ is acyclic, since $(r^*,\hat j)$ is a sink in $\hat{G}$. Further, $\hat{G}$ is maximal since changing only one message label for the decoding choice of only one receiver can alter the size of MAIS by at most one. The claim then follows since $|\hat{G}| =|\texttt{MAIS}(G_{\hat{\mathcal{D}}})|=s+t-i+1$.




\item[3d] \textit{Partially covered group $\mathcal G^*$ is the first layer, and all nodes whose message labels are in $\mathcal{G}^*$ are sink nodes.} As in previous cases, the $t$ sink nodes must have the same receiver label, say $r^*$. Let $\mathcal{G}_{i_1}, \ldots, \mathcal{G}_{i_s}$ denote the groups that receiver $r^*$ possesses. Now, one of three situations arises:
\begin{itemize}
\item[(i)] Layer $\mathcal{L}_1$ has only sink nodes. In this case, this layer has $t$ nodes, and since $\frac{t}{m-s}>g-1$, there must be a group, say $\tilde{\mathcal G}$, that is fully present in layer $\mathcal{L}_1$.
\item[(ii)] There is a non-sink node whose message label lies in a group, say $\tilde{\mathcal G}$, that is fully present in layer $\mathcal{L}_1$.
\item[(iii)] There is a non-sink node whose message label belongs to a group partially present in this layer. This situation is subsumed under Case 3b.
\end{itemize}

There are $t\binom{s+1}{s}>s+t-i$ node representations of receivers that have as side information $s$ groups among $\mathcal{G}_{i_1}, \ldots, \mathcal{G}_{i_s}$ and $\tilde{\mathcal{G}}$. Therefore, one such representation $(r,j^*)$ is absent from $G$. Now, consider a label $\hat j$ such that $X_{\hat j}\in \mathcal{G}^*$, but is not a message label in $G$. Note that $\hat j\notin \mathcal{D}(r)$, since if it were, the induced subgraph consisting of the nodes of $G$ along with $(r,\hat j)$ would be a larger acyclic induced subgraph of $G_{\mathcal D}$\footnote{This is because $(r, \hat j)$ must have the same in-neighborhood as the nodes in $\mathcal{L}_1$ with message labels in $\mathcal{G}^*$, due to Property P4 of Lemma~\ref{lem:GraphProperties}. But nodes in $\mathcal{L}_1$ has no incoming edge, and so neither does $(r, \hat j)$.}, contradicting the fact that $G$ is an MAIS of $G_{\mathcal D}$. Therefore, define a new decoding choice $\hat{\mathcal{D}}$ as follows:
    $$\hat{\mathcal{D}}(x) = \left\{\begin{array}{cc} \mathcal{D}(x) & x\neq r\\ \left(\mathcal{D}(r)\setminus\{j^*\}\right)\cup\{\hat j\} & x=r \end{array}\right..$$
    As in other cases, the subgraph $\hat{G}$ induced by vertices already in $G$ along with $(r,\hat j)$ is an acyclic induced subgraph graph of $G_{\hat{\mathcal{D}}}$. As before, $\hat{G}$ is maximal thereby establishing the existence of a decoding choice $\hat{\mathcal{D}}$ with $|\texttt{MAIS}(G_{\hat{\mathcal{D}}})|=s+t-i+1$. 
\end{itemize}

\section*{Proof of Theorem~\ref{thm:lastcriticalpt}}

By repeated use of Lemma~\ref{lem:IncrementByOne}, we argue that if there is a decoding choice $\mathcal{D}$ for which $|\texttt{MAIS}(G_{\mathcal{D}})|<s+t$, there must exist a decoding choice $\mathcal{D}'$ for which $|\texttt{MAIS}(G_{\mathcal{D}'})|=s+t-1$. In this theorem, we will establish that there is no such $\mathcal{D}'$, thereby establishing that for every valid decoding choice $\mathcal{D}$, $|\texttt{MAIS}(G_{\mathcal{D}})|\geq s+t$. The proof is then complete by the achievability result in Theorem~\ref{thm:achievable}.

Let us assume the existence of a decoding choice $\mathcal{D}'$ for which $|\texttt{MAIS}(G_{\mathcal{D}'})|=s+t-1$. From Corollary~\ref{cor:sabsentgroups}, it follows that there are precisely $s$ groups absent from $\texttt{MAIS}(G_{\mathcal{D}'})$; these are precisely the side information of the receiver label of the sink nodes. Let $\mathcal{F}_1,\ldots,\mathcal{F}_s$ denote these $s$ groups. Note that $\texttt{MAIS}(G_{\mathcal{D}'})$ has exactly $s+t-1=g(m-s)-1$ distinct message labels. So there is precisely one group partially present in $\texttt{MAIS}(G_{\mathcal{D}'})$. Let $\mathcal{F}_{s+1}$ denote this group. Note that $g-1$ message indices from $\mathcal{F}_{s+1}$ are present as message labels in $\texttt{MAIS}(G_{\mathcal{D}'})$. Let the remaining $m-s-1$ groups be denoted by $\mathcal F_{s+2},\ldots,\mathcal F_{m}.$ By hypothesis,
$$s=s+t-t <g(m-s)-(g-1)(m-s)=(m-s).$$
Therefore, $2s+1\leq m$. Construct an $(s+1)\times (2s+1)$ binary matrix $B$ that is tied to $\mathcal{D}'$ as follows:
\begin{itemize}
    \item For $i\in\llbracket s+1\rrbracket$, row $i$ corresponds to receiver with side information $\mathcal{F}_1,\ldots,\mathcal{F}_{i-1}, \mathcal{F}_{i+1},\ldots, \mathcal{F}_{s+1}$ (i.e., the first $s+1$ groups but $\mathcal F_i$). We let $B_{i,j}=1$ and $B_{i,i}=0$ for  $i\in\llbracket s+1\rrbracket$ and $j\in\llbracket s+1\rrbracket\setminus\{i\}$.
    \item For $i\in\llbracket s+1\rrbracket$, and $j=s+1,\ldots, 2s+1$, we set $B_{i,j}=1$ if the receiver corresponding to this row decodes all messages in group $\mathcal F_{j}$ by using its decoding function or by repeatedly emulating other receivers.
\end{itemize}
Now, an application of Lemma~\ref{lem:lemma4liurestated} to the last $s$ columns of $B$ yields a non-empty subset $P\subseteq \llbracket s+1\rrbracket$ of rows of $B$ with the following property:
\begin{itemize}
    \item $|\mathcal{J}|=|P|-1$, where $\mathcal J = \{j\in\{s+2,\ldots,2s+1\}: B_{i,j}=1 \textrm{ for all } i\in P\}$, i.e., $\mathcal J$ collects the indices of groups $\mathcal{F}_{s+2},\ldots, \mathcal{F}_{2s+1}$ that can be fully decoded by each receiver corresponding to row $i\in P$.
\end{itemize}
Now consider the receiver $r^*$ with side information $\left\{\mathcal{F}_k: k\in  \mathcal{J}\cup \big(\llbracket s+1\rrbracket\setminus P\big)\right\}.$ Clearly, this is a bona fide receiver since it has $s+1-|P|+|P|-1=s$ groups as side information. Now, consider the following facts and reasoning:
\begin{itemize}
   \item[F5] By construction, the receiver corresponding to the row $i\in P$ decodes all messages in group $\mathcal{F}_k$, $k\in \mathcal{J}$, and therefore can emulate $r^*$. 
   \item[F6]  $r^*$ decodes at most $(g-1)$ messages from each group $\mathcal{F}_k$ where $k\in\{s+2,\ldots,2s+1\}\setminus\mathcal{J}$. If $r^*$ decodes all messages of any such group by emulating, then by fact F1,  every receiver corresponding to the row $i\in P$ will have also decoded this group, which would necessitate that the index of this group be present in $\mathcal{J}$, a contradiction. This is at most $(g-1)(s+1-|P|)$ messages. 
   \item[F7] Among the other groups, i.e.,  $\mathcal{F}_k$ for $k\in\{2s+2,\ldots,m\}$, $r^*$ can decode at most $g(m-(2s+2)+1)= g(m-2s-1)$ messages.
\end{itemize} 
Combining facts F6 and F7, we see that $r^*$ can decode at most
\begin{align*}
&\hspace{-10mm}(g-1)(s+1-|P|)+g(m-2s-1)\\ &=g(m-s)-s-1-(g-1)|P|\\
&=t-1-(g-1)|P|
\end{align*}
messages from $\mathcal{F}_k$, $k\in\{s+2,\ldots,m\}\setminus\mathcal J$. Since $r^*$ must decode at least $t$ messages, $r^*$ must decode $(g-1)|P|+1$ messages from the groups $\mathcal F_k$, $k\in P$. Note that these $|P|$ groups are precisely those missing from its side information. Thus, $r^*$ must decode, on average, $(g-1)+\frac{1}{|P|}$ messages per group from groups $\mathcal F_k$, $k\in P$. Thus, $r^*$ must decode all messages from a group, say $\mathcal F_{i^*}$ with $i^*\in P$. Now, consider the receiver $\hat r$ that has as side information $\mathcal F_1,\ldots, \mathcal F_{i^*-1}, \mathcal F_{i^*+1}, \ldots, \mathcal F_{s+1}$; this receiver corresponds to row $i^*\in P$. By fact F5, $\hat r$ can decode $\mathcal{F}_{i^*}$ by emulating $r^*$. Now let generalized decoding choice $\mathcal{D}'_{\textrm{app}}$ be defined by 
\begin{align*}
\mathcal{D}'_{\textrm{app}}(x) = \left\{\begin{array}{cc} \mathcal{D}'(x) & x\neq \hat r\\ \mathcal{D}'(x)\cup\{j: j\in \mathcal{F}_{i^*}\} & x=\hat r \end{array}\right.. 
\end{align*}
Consider an induced subgraph of $\mathcal{D}'_{\textrm{app}}$ defined as follows: 
\begin{itemize}
\item[1.] Start with $\texttt{MAIS}(G_{\mathcal{D}'})$ that has $s+t-1$ nodes.
\item[2.] Identify $g-1$ nodes from this subgraph whose message labels belong to the message indices in the partially present group $\mathcal{F}_{s+1}$. Delete these $g-1$ nodes and their incident edges.
\item[3.] Add $g$ nodes $(\hat r, j)$, $j\in\mathcal{F}_{i^*}$ to the subgraph. Add any incoming/outgoing edges to make this graph an induced subgraph of $G_{\mathcal{D}'_{\textrm{app}}}$. 
\end{itemize}
Note that the newly added nodes are sinks in the resultant graph because the message labels originally did not contain any labels that were indices of messages in $\mathcal{F}_1,\ldots, \mathcal{F}_{s}$, and those that were indices of messages in $\mathcal{F}_{s+1}$ were deleted in step 2. Hence, the modification does not induce any cycles. It then follows that the resultant graph, which has $s+t$ nodes is an acyclic induced subgraph of $G_{\mathcal{D}'_{\textrm{app}}}$. Thus, it follows from Corollary~\ref{cor:appendmessages} that $|\texttt{MAIS}(\mathcal{D}')|=|\texttt{MAIS}(\mathcal{D}'_{\textrm{app}})|=s+t$, which is a contradiction. Therefore, no decoding choice $\mathcal D'$ has $|\texttt{MAIS}(\mathcal{D}')|=s+t-1$. Therefore, we infer that
\begin{align*}
R^*(m,s,g,t) \geq \min_{\mathcal{D}} |\texttt{MAIS}(\mathcal{D})|\geq s+t = g(m-s).
\end{align*}
The achievability follows from Theorem~\ref{thm:achievable}. 

\section*{Proof of Theorem~\ref{thm:belowlastcriticalpt}}
   Just as in Theorem~\ref{thm:lastcriticalpt}, the goal here is to show that there is no decoding choice $\mathcal{D}$ such that $|\texttt{MAIS}(\mathcal{D})|= s+t-1$. Let us therefore assume $\mathcal{D}$ is such that $|\texttt{MAIS}(\mathcal{D})|= s+t-1$. From Corollary~\ref{cor:sabsentgroups} it follows that there are precisely $s$ groups absent from $\texttt{MAIS}(G_{\mathcal{D}})$; these are precisely the side information of the receiver label of the sink nodes. Let $\mathcal{A}_1,\ldots,\mathcal{A}_s$ denote these $s$ groups. It then follows that there are $m-s$ partially or fully present groups. Since $|\texttt{MAIS}(\mathcal{D})|= s+t-1$, there must be at least one partially present group. Let's call this $\mathcal{P}_{1}$. Since $s+t-1\geq s+(g-1)(m-s)$, at least $s$ groups must be fully present in $\texttt{MAIS}(G_{\mathcal{D}})$. Let $\mathcal{F}_1,\ldots, \mathcal{F}_s$ denote these groups. 
Let the remaining $m-2s-1$ groups be $\mathcal{P}_2,\ldots, \mathcal{P}_{m-2s}$. Let $y_i$, $i=1,\ldots,  m-2s$ denote the number of message indices in group $\mathcal{P}_i$ that are missing from the message labels in $\texttt{MAIS}(G_{\mathcal{D}})$. By choice, $y_1>0$, and 
    \begin{align}
        \sum_{k=1}^{m-2s}y_k=g(m-s)-(s+t-i).\label{eqn:ysum}
    \end{align}

    Now, just as in Theorem~\ref{thm:lastcriticalpt}, let us define an $(s+1)\times (2s+1)$ binary matrix $B$ that is tied to $\mathcal{D}$ as follows:
\begin{itemize}
    \item For $i\in\llbracket s\rrbracket$, row $i$ corresponds to receiver with side information $\mathcal{A}_1,\ldots,\mathcal{A}_{i-1}, \mathcal{A}_{i+1},\ldots, \mathcal{A}_s$ and $\mathcal{P}_1$. We let $B_{i,j}=1$ and $B_{i,i}=0$ for  $i\in\llbracket s\rrbracket$ and $j\in\llbracket s+1\rrbracket \setminus\{i\}$.
    \item Row $s+1$ corresponds to receiver with side information $\mathcal{A}_1,\ldots,\mathcal{A}_s$. We let $B_{s+1,j}=1$ iff $j\in\llbracket s\rrbracket$.
    \item For $i\in\llbracket s+1\rrbracket$, and $j>s+1$, we set $B_{i,j}=1$ if the receiver corresponding to this row decodes all messages in group $\mathcal F_{j-(s+1)}$ by using its decoding function or by repeatedly emulating other receivers.
\end{itemize}
    Now, an application of Lemma~\ref{lem:lemma4liurestated} to the last $s$ columns of $B$ will yield a non-empty subset $P\subseteq \llbracket s+1\rrbracket$ of rows of $B$ with the following property:
\begin{itemize}
    \item $|\mathcal{J}|=|P|-1$, where $\mathcal J = \{j\in\llbracket s\rrbracket: B_{i,j+s+1}=1 \textrm{ for all } i\in P\}$, i.e., $\mathcal J$ collects the indices of groups $\mathcal{F}_{1},\ldots, \mathcal{F}_{s}$ that can be fully decoded by each receiver corresponding to row $i\in P$.
\end{itemize}

Now, if $s+1\in P$, define $r^*$ to be the receiver with side information $\mathcal A_i$ for $i\in\llbracket s\rrbracket\setminus P$, and $\mathcal F_j$ for $j\in \mathcal{J}$; and if $s+1\notin P$, define $r^*$ to be the receiver with side information $\mathcal{P}_1$, $\mathcal A_i$ for $i\in\llbracket s\rrbracket\setminus P$, and $\mathcal F_j$ for $j\in \mathcal{J}$. Clearly, this is a bona fide receiver since it has $s+1-|P|+|P|-1=s$ groups as side information. Now, consider the following:
\begin{itemize}
   \item[F8] By construction, every receiver corresponding to the row $i\in P$ recovers all messages in the groups in $\mathcal{J}$, and therefore can emulate $r^*$. 
   \item[F9]  $r^*$ decodes at most $(g-1)$ messages from each group  $\mathcal F_k$ for $ k\in\llbracket s\rrbracket\setminus\mathcal{J}$. If $r^*$ decodes all messages of any such group by emulating, then by fact F8,  every receiver corresponding to the row $i\in P$ will have also decoded this group, which would necessitate that the index of this group be present in $\mathcal{J}$, a contradiction. So $r^*$ can decode at most $(g-1)$ messages from each group  $\mathcal F_k$ for $ k\in\llbracket s\rrbracket\setminus\mathcal{J}$. This is at most $(g-1)(s+1-|P|)$ messages. 
\end{itemize} 
We now divide the proof into two cases.
\begin{itemize}
\item Case 1: $s+1\in P$. In this case, $r^*$ has $\mathcal{A}_i$ for $i\in P$, and $\mathcal F_j$ for $j\in \mathcal{J}$ as side information. Let us count the number of messages $r^*$ can decode from messages in $\mathcal{F}_1,\ldots, \mathcal{F}_s$ and only those message indices in $\mathcal{P}_1,\ldots, \mathcal{P}_{m-2s}$ that are present in $|\texttt{MAIS}(\mathcal{D})|$. The former computation is already given in fact F9. The latter is at most $\sum_{k=1}^{m-2s} (g-y_i).$ Together, the sum comes to
\begin{align*}
 (s + 1 - |P|)(g - 1) &+ \sum_{i = 1}^{m - 2s}(g - y_i) \\
 &\stackrel{\eqref{eqn:ysum}}{=} t-\Big( 1+(|P|-1)(g - 1)\Big).
\end{align*}
Since $r^*$ must decode $t$ messages, $r^*$ must decode $1+(|P|-1)(g - 1)\geq 1$ messages from $\mathcal A_i$ for $i\in P\cap \llbracket s\rrbracket$ or from the message indices in $\mathcal{P}_1,\ldots, \mathcal{P}_{m-2s}$ that are missing in $\texttt{MAIS}(\mathcal{D})$. Now, two cases arise:
\begin{itemize}
    \item[1a:] $r^*$ decodes a message $X_{\hat j}$ that is in $\mathcal{P}_1,\ldots, \mathcal{P}_{m-2s}$ but is missing in $\texttt{MAIS}(\mathcal{D})$. Since in this setting $s+1\in P$, by fact F8, the receiver  $r_{\textrm{sink}}$ with side information $\mathcal{A}_1,\ldots, \mathcal{A}_s$ can mimic $r^*$ to also decode $\hat j$. Then, we can append $\hat j$ to $\mathcal{D}(r_{\textrm{sink}})$ to obtain a generalized decoding choice $D^*$. The induced subgraph of $G_{\mathcal{D}^*}$ consisting of the nodes of $\texttt{MAIS}(\mathcal{D})$ along with $(r_{\textrm{sink}}, \hat j)$ has a size of $s+t$ and is also acyclic. Therefore, it follows that  $|\texttt{MAIS}(\mathcal{D}^*)|\geq s+t$. However, by Corollary~\ref{cor:appendmessages}, $s+t\leq |\texttt{MAIS}(\mathcal{D})|= |\texttt{MAIS}(\mathcal{D}^*)|=s+t-1$, which is a contradiction. 
    \item[1b:] $r^*$ decodes no messages in $\mathcal{P}_1,\ldots, \mathcal{P}_{m-2s}$ that are missing in $\texttt{MAIS}(\mathcal{D})$. In this case, $r^*$ must decode $1+(|P|-1)(g - 1)$ messages from the $|P|-1$ groups not in its side information, i.e., from $\mathcal A_i$ for $i\in P\cap \llbracket s\rrbracket$. This would mean that it must decode at least one group, say $\mathcal A_{i^*}$ fully by emulating other nodes. By invoking F8, we can argue that receiver $\hat r$ with side information $\mathcal A_1,\ldots, \mathcal A_{i^*-1},\mathcal A_{i^*+1}, \mathcal{A}_s, \mathcal{P}_1$ can mimic $r^*$ and decode messages in $\mathcal A_{i^*}$ by successive emulation. In this case, let generalized decoding choice $\mathcal{D}^*$ append to $\mathcal D$ the $g$ new messages of $\mathcal A_{i^*}$ that $\hat r$ decodes. Now, consider the induced subgraph of $G_{\mathcal{D}^*}$ obtained by deleting from $\texttt{MAIS}(\mathcal{D})$ at most $g-1$ nodes whose message labels belong to $\mathcal{P}_1$, and then adding $g$ nodes $\{(\hat r, k): X_k\in \mathcal A_{i^*}\}$  (and any incident edges). This subgraph is acyclic and has a size of at least $s+t$. Again, by Corollary~\ref{cor:appendmessages}, $s+t\leq |\texttt{MAIS}(\mathcal{D})|= |\texttt{MAIS}(\mathcal{D}^*)|=s+t-1$, which is a contradiction.
\end{itemize}
\item Case 2: $s+1\notin P$. In this case, $r^*$ has $\mathcal{P}_1$, $\mathcal{A}_i$ for $i\in P$, and $\mathcal F_j$ for $j\in \mathcal{J}$ as side information. As before, let us count the number of messages $r^*$ can decode from  messages in $\mathcal{F}_1,\ldots, \mathcal{F}_s$ and only those message indices in $\mathcal{P}_2,\ldots, \mathcal{P}_{m-2s}$ that are present in $\texttt{MAIS}(\mathcal{D})$. The former computation is already given in fact F9. The latter is given by $\sum_{k=2}^{m-2s} (g-y_i).$ Together, the sum comes to
\begin{align*}
 (s + 1 - |P|)(g - 1) &+ \sum_{i = 2}^{m - 2s}(g - y_i) \\
 &\stackrel{\eqref{eqn:ysum}}{=} t-\big((g - 1)|P|+2-y_1\big).
\end{align*}
Thus, $r^*$ has to decode $(g - 1)|P|+2-y_1$ messages from $\mathcal A_i$ for $i\in P$ or from the message indices in $\mathcal{P}_1,\ldots, \mathcal{P}_{m-2s}$ that are missing in $\texttt{MAIS}(\mathcal{D})$. Suppose that $r^*$ decodes $x$ messages whose indices are in $\mathcal{P}_1,\ldots, \mathcal{P}_{m-2s}$, but are missing in $\texttt{MAIS}(\mathcal{D})$. Then, $r^*$ decodes $(g - 1)|P|+2-y_1-x$ messages from $\mathcal A_i$, $i\in P$. Thus, there must exist a group $\mathcal A_{i^*}$, $i^*\in P$ such that $r^*$ decodes at least $\frac{(g - 1)|P|+2-y_1-x}{|P|}$ messages in $\mathcal A_{i^*}$. 

From F8, we see that the receiver $\hat r$ with side information $\mathcal A_1,\ldots, \mathcal A_{i^*-1},\mathcal A_{i^*+1},\ldots \mathcal{A}_s, \mathcal{P}_1$ emulates $r^*$ to decode 
\begin{align*}
\frac{(g - 1)|P|+2-y_1-x}{|P|}+x > g -y_1  
\end{align*}
since $(|P| - 1)(x + y_1) - |P| + 2 > 0$ because $x+y_1>0$. Thus, we see that the receiver $\hat r$ decodes at least $g-y_1+1$ messages, say $X_{k_1},\ldots, X_{k_{g-y_1+1}}$, that are not part of $\mathcal{D}(\hat r)$. Consider the generalized decoding choice $\mathcal{D}^*$ by appending to $\mathcal D$ the  new messages that $\hat r$ decodes. Next, consider the induced subgraph of $G_{\mathcal{D}^*}$ obtained by deleting from $\texttt{MAIS}(\mathcal{D})$ $g-y_1$ nodes whose message labels belong to $\mathcal{P}_1$, and then adding $g-y_1+1$ nodes $\{(\hat r, k_i): 1\leq i\leq g-y_1+1\}$ (and any incident edges). This subgraph is acyclic and has a size of $s+t$. Again, by Corollary~\ref{cor:appendmessages}, it follows that $s+t\leq |\texttt{MAIS}(\mathcal{D})|= |\texttt{MAIS}(\mathcal{D}^*)|=s+t-1$, which is a contradiction.
\end{itemize}
Thus, in all cases, we reach a contradiction, which implies that our assumption of the existence of a decoding choice $D$ with $|\texttt{MAIS}(\mathcal{D})|=s+t-1$ is untenable. Therefore, 
\begin{align*}
R^*(m,s,g,t) \geq \min_{\mathcal{D}} |\texttt{MAIS}(\mathcal{D})|\geq s+t .
\end{align*}
The proof is complete due to the achievability in Theorem~\ref{thm:achievable}. 

\section*{Proof of Theorem~\ref{thm:abovelastcriticalpt}}
As in the proof of Proposition~7 of \cite{liu2019tight}, we choose $\alpha = s+t-g(m-s)$. Note that $0< \alpha \leq s$. Suppose we look at a subproblem of the PICOD problem where we focus on delivering $t$ messages to receivers that have $\mathcal{G}_1,\ldots, \mathcal{G}_{\alpha}$ and any $s-\alpha$ other groups as side information. This subproblem has $\binom{m-\alpha}{s-\alpha}$ receivers. Note that a lower bound on the optimal rate of this subproblem is also a lower rate for $R^*(m,s,g,t).$ Further, any code for the subproblem need only focus on the $m-\alpha$ groups that are not common to all receivers. Hence, this is effectively an $(m-\alpha, s-\alpha, g, t)$ PICOD problem. But note that $(s-\alpha)+t = g((m-\alpha)-(s-\alpha))$ and $t>(g-1)((m-\alpha)-(s-\alpha))$. Hence, from Theorem~\ref{thm:lastcriticalpt}, we see that a lower bound to this sub-problem, and hence a lower bound for $R^*(m,s,g,t)$ is given by $g(m-s)$. The upper bound follows from the achievability in Theorem~\ref{thm:achievable}. 

\section*{Proof of Theorem~\ref{thm:restrictedconverse}}
From Theorem~\ref{thm:achievable}, it is clear that by sending $f-1$ MDS sequentially at a rate of $m-s$ each followed by either an uncoded transmission of $s+(t-(f-1)(m-s))$ messages (when $s+t<f(m-s)$) or another MDS code at a rate of $m-s$ (when $s+t\geq f(m-s)$) suffices to meet all receiver demands. Hence, the optimal broadcast rate among codes from $\mathcal{C}_{m,s,g,t}$ is no more than $f(m-s)$.

Consider an asymptotically vanishing error (lossless) variant of the PICOD problem where the goal is to jointly encode $n$ i.i.d. copies of the $mg$ messages to deliver the $n$ realizations of some $t$ messages to each receiver. By a simple information-theoretic argument, we can establish that the optimal rate (normalized by message size) for this relaxed variant when restricted only to broadcast codes is identical to the maximum number of messages any receiver recovers. Additionally, from Theorem~2 of \cite{langberg2011network}, it follows that the optimal rate of the relaxed variant of the PICOD problem and optimal rate among codes from $\mathcal{C}_{m,s,g,t}$ for the zero-error PICOD formulation in this work are identical. Thus, the optimal rate among codes from $\mathcal{C}_{m,s,g,t}$ for the PICOD formulation in this work is identical to the maximum number of messages communicated to any receiver. 

Now, let $C$ be a code of optimal rate from this family. For this code, let the encoder communicate $0\leq a_i\leq g$ messages from the group $\mathcal G_i$, $i\in\llbracket m \rrbracket$. Without loss of generality, we can relabel the groups so that $a_1\geq a_2\geq \cdots \geq a_m$. Consider the receiver that has the last $s$ groups and therefore decodes the most messages among all receivers. Then, this receiver decodes $a_{1}+a_2+\cdots+a_{m-s}$ messages. Hence $C$'s broadcast rate must equal $\sum_{j=1}^{m-s} a_j$. From the achievability argument, we see that $ \sum_{i=1}^{m-s}a_i\leq R'(m,s,g,t)\leq f(m-s)$. Further, the least number of messages that any receiver decodes is by the receiver that has the first $s$ groups (after relabeling) as side information, and therefore, $\sum_{i=s+1}^m a_i \geq t$.

 If $a_1-a_m>1$, then consider another assignment that conveys $b_1=a_1-1$, $b_i=a_i$ for $1<i<m$, and $b_m=a_m+1$ messages from each group. The least number of messages that any receiver decodes under $C'$ is the sum of the $m-s$ smallest $b$'s; this sum is either $a_s,\ldots,a_{m-1}$ or $a_{s+1},\ldots, a_{m-1}, a_m+1$. In any case, the sum is larger than that of $C$. Therefore, every receiver will be satisfied under $C'$ as well. Further, the most number of messages that any receiver decodes is given by the sum of $m-s$ largest $b$'s, which is  either $a_1-1,a_2,\ldots, a_{m-s}$ or $a_2,\ldots, a_{m-s+1}$. In either case, the sum of $m-s$ largest $b$'s is at most $\sum_{i=1}^{m-s} a_i \geq t$. We can relabel the groups to sort the $b$'s in non-increasing order, and repeat this approach of balancing where we attempt to reduce/increase the maximum/minimum number of messages conveyed from a group, respectively. This balancing will end with a configuration that conveys $\alpha_1\geq \alpha_2\geq \cdots \geq \alpha_m$ messages from each group (after relabeling) where $\alpha_1-\alpha_m\leq 1$. Thus, we see that
 $$\sum_{i=1}^{m-s}\alpha_i\leq \sum_{i=1}^{m-s}a_i\leq R'(m,s,g,t)\leq f(m-s).$$
 Thus it follows that $\alpha_i\in\{f-1,f\}$ for each $i\in\llbracket m\rrbracket$. 
 Since no more than $f$ messages are conveyed per group, we can remove/ignore $g-f$ messages per group, and consider an $f$-group complete-$\{s\}$ PICOD($t$) problem with $m$ groups. Since $t$ now meets the hypothesis of Theorems~\ref{thm:belowlastcriticalpt} and \ref{thm:abovelastcriticalpt}, we see that the lowest rate for any code (broadcast or not) is at least $\min\{s+t, f(m-s)\}$. Therefore, it follows that 
 $$\min\{s+t, f(m-s)\}\leq \sum_{j=1}^{m-s}\alpha_i \leq R'(m,s,g,t)$$
 The achievability follows from Theorem~\ref{thm:achievable} by noting that the code construction results in a broadcast code.

\section*{A Technical Result}
The following is an equivalent formulation of Lemma~4 of \cite{liu2019tight} stated here without proof. It is invoked in the proofs of Theorems~\ref{thm:lastcriticalpt} and \ref{thm:belowlastcriticalpt}.
\begin{lemma}
\label{lem:lemma4liurestated}
Let $B$ be an $(s+1)\times s$ binary matrix for some $s>0$. Then there exists a non-empty subset $P\subseteq \llbracket s+1 \rrbracket$ such that $\{j: B_{i,j}=1 \textrm{ for every } i \in P\}= |P|-1$.
\end{lemma}


\end{document}